\documentclass{article}

\usepackage{amsmath,amstext,amsthm,amssymb,amsfonts}
\usepackage{graphicx,psfrag}
\usepackage[top=1in,bottom=1in,left=1.25in,right=1.25in]{geometry}
\usepackage{caption}
\usepackage{subcaption}

\newtheorem{theorem}{Theorem}

\newtheorem{prop}{Proposition}

\DeclareMathOperator{\cost}{cost}
\DeclareMathOperator{\eqf}{eqf}

\DeclareMathOperator{\poly}{poly}

\begin{document}

\title{Computing Optimal Tolls in Routing Games without Knowing the Latency Functions}
\author{Siddharth Barman\footnote{California Institute of Technology. email: barman@caltech.edu} \qquad Umang Bhaskar\footnote{University of Waterloo. email: umang@caltech.edu} \qquad Chaitanya Swamy\footnote{University of Waterloo. email: cswamy@uwaterloo.ca}}

\maketitle

{\small \paragraph{Abstract.}
We consider the following question: in a nonatomic routing game, can the tolls that induce the minimum latency flow be computed without knowing the latency functions? Since the latency functions are unknown, we assume we are given an oracle that has access to the underlying routing game. A query to the oracle consists of tolls on edges, and the response is the resulting equilibrium flow. We show that in this model, it is impossible to obtain optimal tolls. However, if we augment the oracle so that it returns the total latency of the equilibrium flow induced by the tolls in addition to the flow itself, then the required tolls can be computed with a polynomial number of queries.}

\vspace{2em}

\section{Introduction and Preliminaries}
 
Let $\varGamma = (G=(V,E),l,K=(d_i,s_i,t_i)_{i \in [k]})$ be a multicommodity nonatomic routing game on a directed graph $G=(V,E)$, where $l = (l_e)_{e \in E}$ are the latency functions on the edges, and there are $k$ commodities. Commodity $i$ has source $s_i$, sink $t_i$, and total demand $d_i$. As is standard, $m := |E|$ and $n := |V|$. 

A multicommodity flow $f = (f^i)_{i \in [k]}$ is feasible if each flow $f^i$ is an $s_i$-$t_i$ flow of value $d_i$. A feasible flow $f$ is a Wardrop equilibrium (or simply an equilibrium) if for every commodity $i$ and all $s_i$-$t_i$ paths $P$, $Q$ with $f_e^i > 0$ on every edge $e \in P$,

\[
\sum_{e \in P} l_e(f_e) \le \sum_{e \in Q} l_e (f_e) \, .
\]

For latency functions $l$ in game $\varGamma$, we use $\eqf(l)$ to denote the equilibrium flow. For any flow $f$, we define $\cost(f,l) = \sum_{e \in E} f_e l_e(f_e)$ as the total latency of flow $f$. The optimal flow for a game is the feasible flow that minimizes the total latency.

Let $\tau = (\tau_e)_{e \in E} \in \mathbb{R}^m_+$ be a vector of non-negative tolls on edges. Then a feasible flow $f$ is a Wardrop equilibrium (or simply an equilibrium) with tolls $\tau$ if for every commodity $i$ and all $s_i$-$t_i$ paths $P$, $Q$ with $f_e^i > 0$ on every edge $e \in P$, 

\[
\sum_{e \in P} l_e(f_e) + \tau_e \le \sum_{e \in Q} l_e (f_e) + \tau_e \, .
\]

\noindent We use $\eqf(l,\tau)$ to denote the equilibrium flow with tolls $\tau$ on the edges, and say that tolls $\tau$ induce the flow $\eqf(l, \tau)$. 

We study a query model introduced in~\cite{BhaskarLSS14}, as well as its extension, for routing games. In this model, there is a routing game $\varGamma$, the latency functions $l_e^*$ of which are unknown. Instead, we are given oracle access to the routing game. Each query to the oracle consists of tolls $\tau = (\tau_e)_{e \in E}$ on the edges, and the response is the equilibrium flow $\eqf(l^*, \tau)$. For our results, we restrict the latency functions to \emph{standard degree-$r$ polynomials} as defined in~\cite{BhaskarLSS14}. Then $\mathcal{I}$, $U$ and $K$ are defined accordingly. In particular, $\mathcal{I}$ is the input size of the routing game, $\log U = \poly(\mathcal{I})$, and $K := K(r)  = \poly(U, \sum_i d_i)$. 

Results similar to those in Section~\ref{sec:positive} were also independently obtained by Roth et al.~\cite{RothUW15}.

\section{Computing Optimal Tolls}

\subsection{Optimal tolls from observations of equilibrium flow}

\begin{figure}[!h]
\centering
\begin{subfigure}{0.4\textwidth}
\centering
\psfrag{l1}{$x$}
\psfrag{l2}{$0$}
\includegraphics[width=0.4\linewidth]{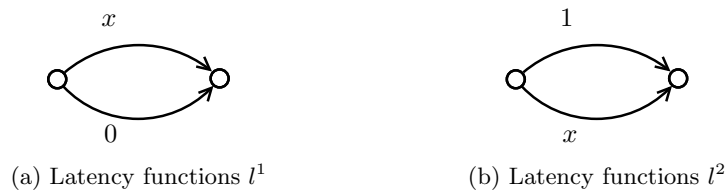}
\caption{Latency functions $l^1$}
\end{subfigure}%
\begin{subfigure}{0.4\textwidth}
\centering
\psfrag{l1}{$1$}
\psfrag{l2}{$x$}
\includegraphics[width=0.4\linewidth]{parallel-twolinks}
\caption{Latency functions $l^2$}
\end{subfigure}
\caption{Latency functions $l^1$ and $l^2$ have the same equilibrium flow for any edge tolls. Demand $d=1$ in the example.}
\label{fig:equivalent}
\end{figure}

We first show that if in response to a query the oracle returns only the flow at equilibrium, then it is impossible to obtain the optimal tolls. For this, we construct two single-commodity routing games $\varGamma^1$ and $\varGamma^2$ with demand $d=1$ (Figure~\ref{fig:equivalent}) on parallel edges with latency functions $l^1$ and $l^2$ respectively that have the same equilibrium flow for any tolls, and hence cannot be distinguished in the query model.\footnote{This is the query model in~\cite{BhaskarLSS14}, and shows that in this model it is necessary to be given a particular target flow.} However, the optimal flows and optimal tolls are different for the two routing games.

It can be easily verified that for any tolls $\tau$, $\eqf(l^1, \tau) = \eqf(l^2, \tau)$. However, for $l^1$, the equilibrium flow is the optimal flow, while for $l^2$, the optimal flow is $(1/2, 1/2)$.

\subsection{Optimal tolls from observations of equilibrium flow and total latency}
\label{sec:positive}

Given the previous impossibility result, an obvious question is whether the optimal tolls can be obtained with limited additional information from the oracle in response to a query. We now show that if in response to a query, the oracle returns the flow at equilibrium \emph{as well as the total latency at equilibrium}, the optimal tolls can in fact be obtained with a polynomial number of queries. 

The starting point for our result is the observation that the total latency is a strictly convex function of the flows. By this observation, if we could obtain the total latency and its gradient for any flow, then standard gradient-descent algorithms would suffice to obtain the minimum-latency flow. By the ellipsoid-based algorithm in~\cite{BhaskarLSS14}, given any acyclic target flow, we can obtain the tolls required to enforce the flow with a polynomial number of queries. Since we assume we additionally obtain from the oracle the total latency of the equilibrium flow, the first of the requirements can be satisfied: for any flow, we can obtain the tolls required to enforce it, as well as its total latency. Additionally, if we are given the optimal flow, we can obtain the tolls required to enforce it as well.

\begin{theorem}[\cite{BhaskarLSS14}] 
Let $f^*$ be a target acyclic multicommodity flow and $\delta > 0$. Let $\epsilon = \frac{\delta^2}{Kmk \sum_i d_i}$. Then, in time $\poly\left(\mathcal{I},\log(\frac{1}{\delta})\right)$ and using $\poly\left(\mathcal{I},\log(\frac{1}{\delta})\right)$ $\epsilon$-oracle queries, we can compute tolls $\tau$ such that $\Vert f(l^*,\tau) - f^*\Vert_\infty \le 2 \delta$ or determine that no such tolls exist.
\label{thm:tolls}
\end{theorem}

\noindent It follows from the theorem that for any $\delta > 0$, we can obtain in a polynomial number of queries tolls $\tau$ so that $\Vert f(l^*,\tau) - f^*\Vert_\infty \le \delta / (2 m K^2)$, and hence $\cost(f(l^*,\tau), l^*) - \cost(f^*, l^*) \le \delta$.

In order to obtain the optimal tolls, instead of trying to obtain the gradient, we utilize the \emph{zero-order} convex programming techniques of~\cite[Chapter 9]{NemirovskyY83}, based on the ellipsoid algorithm. These algorithms return an approximate minimizer of the convex function, but require only a zero-order oracle for minimizing a convex function: an oracle which returns only the value of the function at the given point. We present the results from~\cite[Chapter 9]{NemirovskyY83} for opimization using a zero-order oracle, and then describe its use in our algorithm.

\paragraph*{Optimization with a zero-order oracle.} We modify (and somewhat simplify) the notation from~\cite{NemirovskyY83} in order to avoid conflicts with notation already used. Let $C$ be a convex, closed, bounded body in $\mathbb{R}^N$. We require that there exist ellipsoids $\bar{W}^0$ and $W^0$ so that $\bar{W}^0 \supset C \supset W^0$ and $|\bar{W}^0| \le N^N |W^0|$. This can be enforced by standard techniques (see~\cite[Chapter 4]{NemirovskyY83}). Let $f_0(x)$, $\ldots$, $f_M(x)$ be convex and continuous functions defined on $C$. We also have a zero-order $\delta$-oracle $\mathbb{\psi} = (\psi_0, \ldots, \psi_M)$ so that for $x \in C$, $| \psi_j(x) - f_j(x) | \le \delta$ for $j = 0, \ldots, M$. The class of problems

\[
\min_{x \in C} f_0(x) ~ | ~ f_j(x) \le 0, ~ j \in [M]
\]

\noindent is denoted $\mathcal{C}^\delta(C, \mathbb{R}^N,M)$. Define 

\[
r_j(f) := \left \{ \begin{array}{ll}
	\sup_C f_0 - \inf_C f_0 & \mbox{ if $j=0$,} \\
	\max \{0, \, \sup_C f_i\} & \mbox{ if $j \in [M]$} \, .
\end{array} \right.
\]

\noindent Further, for $\epsilon > 0$ define 

\[
\nu(f,\epsilon) := \min_j \frac{\epsilon}{r_j(f)} \, .
\]

\begin{theorem}[\cite{NemirovskyY83}]
There exists a polynomial $P(N)$ so that if $\epsilon \ge P(N) \delta$ then any problem from the class $\mathcal{C}^\delta(C, \mathbb{R}^N, M)$ can be solved to within absolute error $2\epsilon$ in time $\displaystyle O\left(N^7 \ln N \ln \frac{4N^2}{\nu(f, \epsilon)}\right)$.
\label{thm:ny}
\end{theorem}

For our algorithm, we define $C$ to be the set of feasible multicommodity flows. Then $N = mk$, $M = 0$. Define $f_0(f)$ to be the total latency of flow $f$. By definition of $K$, $r_0(f) \le mK (\sum_i d_i)^2$. Let $\epsilon > 0$ and let $f^*$ be the minimum latency flow. Theorem~\ref{thm:ny} gives the following result in this case.

\begin{prop}
A feasible flow $\hat{f}$ that satisfies $\cost(\hat{f}, l^*) - \cost(f^*,l^*) \le \epsilon$ can be obtained with $O\left(N^7 \log N \log (NK\sum_i d_i/\epsilon)\right)$ queries to a zero-order $\delta$-oracle that returns the total latency of the flow.
\label{prop:optimization}
\end{prop}

The algorithm for obtaining the optimal tolls then follows by combining the algorithm for Proposition~\ref{prop:optimization} with the oracle obtained from Theorem~\ref{thm:tolls}. 

\bibliographystyle{plain}
\bibliography{optimal-tolls}

\end{document}